\documentclass[
 reprint, amsmath,amssymb, aps, superscriptaddress
]{revtex4-2}

\usepackage[letterpaper, total={6.5in, 9in}]{geometry}

\usepackage[braket]{qcircuit}
\usepackage{xcolor}
\usepackage{graphicx}
\usepackage{hyperref}
\usepackage{siunitx}
\usepackage{float}

\usepackage{array}
\newcolumntype{P}[1]{>{\raggedright\arraybackslash}p{#1}} 

\begin{document}

\title{Faster and More Reliable Quantum SWAPs via Native Gates}

\author{Pranav Gokhale}
\affiliation{Super.tech}
\author{Teague Tomesh}
\affiliation{Super.tech}
\affiliation{Princeton University}
\author{Martin Suchara}
\affiliation{Argonne National Laboratory}
\author{Frederic T. Chong}
\affiliation{Super.tech}
\affiliation{University of Chicago}

\date{\today}

\begin{abstract}
Due to the sparse connectivity of superconducting quantum computers, qubit communication via SWAP gates accounts for the vast majority of overhead in quantum programs. We introduce a method for improving the speed and reliability of SWAPs at the level of the superconducting hardware's native gateset. Our method relies on four techniques: 1) SWAP Orientation, 2) Cross-Gate Pulse Cancellation, 3) Commutation through Cross-Resonance, and 4) Cross-Resonance Polarity. Importantly, our Optimized SWAP is bootstrapped from the pre-calibrated gates, and therefore incurs zero calibration overhead. We experimentally evaluate our optimizations with Qiskit Pulse on IBM hardware. Our Optimized SWAP is 11\% faster and 13\% more reliable than the Standard SWAP. We also experimentally validate our optimizations on application-level benchmarks. Due to (a) the multiplicatively compounding gains from improved SWAPs and (b) the frequency of SWAPs, we observe typical improvements in success probability of 10--40\%. The Optimized SWAP is available through the SuperstaQ platform.

\end{abstract}

\maketitle

\section{Introduction}
Superconducting quantum computers typically have sparse qubit connectivity. As a result, operations between distant qubits must be bridged by SWAP gates. To limit the overhead of SWAP gates, researchers have proposed many techniques \cite{javadi2017optimized, siraichi2018qubit, zulehner2018efficient, tannu2019not, li2019tackling, wille2019mapping, cowtan2019qubit, shi2020resource, zhang2020context} for intelligently mapping logical program qubits to physical hardware qubits in order to minimize qubit communication. While qubit mapping does mitigate the core problem, SWAPs still account for an overwhelming fraction of operations on superconducting quantum processors. This fraction approaches 100\% in the limit of larger systems, since the average pairwise distance between qubits increases when we add qubits to typical device topologies. Moreover, light-cone oriented studies \cite{tomesh2020coreset, farhi2020quantum, bravyi2019obstacles} observe that near-term quantum advantage will require a dense logical qubit interaction graph, suggesting that the SWAP overhead is unlikely to diminish.

We address the this bottleneck by improving the fidelity of the SWAP operation itself. Our work relies on lower-level primitives for quantum computers than the typical gate-based based abstraction that programmers typically work with. Instead, we work at the level of \textit{native gates} that capture the natural capabilities of underlying qubits.

We highlight the importance of improving SWAP gates from the perspective of Amdahl's Law \cite{amdahl1967validity}: speeding up a component by a factor $s$ leads to an overall system speedup of $1/(1-p+\frac{p}{s})$, where $p$ is fraction of time spent on the component. Intuitively, Amdahl's Law guides us to consider modest (low $s$) improvements that target bottleneck components (very high $p$)---such as SWAPs. In the quantum setting, the implication is even more consequential: component speedups will lead to \textit{multiplicatively} compounding fidelity gains by avoiding errors due to the limited qubit coherence lifetimes. In this paper, we show that modest gains in the speed and fidelity of qubit communication lead to significant improvements on full applications, since the associated $p$ is almost 100\%.

The remainder of this paper is organized as follows. Section~\ref{sec:background} provides background material, and Section~\ref{sec:swap_orientation} presents a warm-up optimization called SWAP Orientation. Section~\ref{sec:optimized_swap} presents the other three optimizations, culminating in our final Optimized SWAP. Section~\ref{sec:results} shows the results of our Optimized SWAP on IBM Quantum hardware. Section~\ref{sec:conclusion} discusses these results and concludes. Appendix~\ref{app:rb} includes details for our Interleaved Randomized Benchmarking of our Optimized SWAP gates. Appendix~\ref{app:pulse_schedules} has the exact pulse schedules we executed for our Optimized SWAP.

\section{Background} \label{sec:background}

\subsection{Basics of SWAP}
For a classical bit, the standard temporary-variable assisted SWAP suffices: \texttt{temp  $\leftarrow$ \texttt{q\textsubscript{1}}; \texttt{q\textsubscript{1}} $\leftarrow$ \texttt{q\textsubscript{2}}; \texttt{q\textsubscript{2}} $\leftarrow$ temp}. However, as a consequence of the \textit{No-Cloning Theorem} \cite{wootters1982single}, qubits---unlike classical bits---cannot be copied into a temporary register. This restriction forbids the temporary-variable assisted technique for quantum SWAPs.

Quantum SWAPs are instead implemented using the XOR trick for in-place classical SWAPs: \texttt{q\textsubscript{1} $\leftarrow$ q\textsubscript{1} XOR q\textsubscript{2}; q\textsubscript{2} $\leftarrow$ q\textsubscript{2} XOR q\textsubscript{1}; q\textsubscript{1} $\leftarrow$ q\textsubscript{1} XOR q\textsubscript{2}}. The XOR operations can be performed by the CNOT (Controlled-NOT) gate. Thus, a SWAP can be implemented by a sequence of three CNOTs with alternating orientation.

\begin{figure}[h]
$$
\Qcircuit @C=0.7em @R=0.7em {
& \Qswap      & \qw && \raisebox{-2.2em}{=} &&& \Ctrl{1} & \Targ & \Ctrl{1} & \qw  & & \raisebox{-2.2em}{=} &&& \Targ & \Ctrl{1} & \Targ & \qw \\
& \Qswap \qwx & \qw &&   &&& \Targ    & \Ctrl{-1} & \Targ & \qw &&   &&& \Ctrl{-1} & \Targ & \Ctrl{-1} & \qw
}
$$
\caption{SWAP can be implemented as either CNOT-NOTC-CNOT or NOTC-CNOT-NOTC. While the two are symmetric, native gate decomposition reveals that one direction is faster.} \label{fig:basic_swap}
\end{figure}
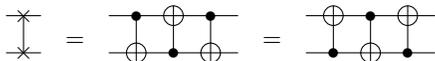

Figure~\ref{fig:basic_swap} depicts this decomposition as an equivalence of three quantum circuits. To disambiguate, we refer to a Controlled-NOT as CNOT when the control is the top qubit or NOTC when the control is on the bottom qubit. As depicted, a SWAP can be implemented as either CNOT-NOTC-CNOT or as NOTC-CNOT-NOTC. While these two decompositions are logically equivalent, they have different implications in terms of speed and fidelity, as we elaborate in Section~\ref{sec:swap_orientation}.

\subsection{Native Gates}
Quantum programming languages are typically structured akin to Hardware Description Languages in classical computing. For example, the Qiskit \cite{Qiskit}, Cirq \cite{cirq}, and Q\# \cite{svore2018q} quantum languages all revolve around quantum circuit primitives that are constructed from a mix of operations ranging from simple ones like $R_x$ gates to complex ones like Quantum Fourier Transforms. However, to actually run on real hardware, these programs must be compiled down to an assembly language. OpenQASM \cite{cross2017open, mckay2018qiskit} is the most widely used assembly \cite{singhalverified}, and it is backed by wide academic study such as LLVM integration \cite{litteken2020updated}, conversion tools \cite{iten2019introduction}, benchmarks \cite{li2020qasmbench}, and formal verification \cite{singhalverified}.

OpenQASM specifies a small set of allowed \textit{basis gates} that every quantum program can be compiled down to. The 2-arity basis gate is the CNOT gate, which is a convenient choice since (a) it has an intuitive meaning and (b) textbook presentations of quantum algorithms are typically written in terms of CNOTs. It is conceptually appealing to think of the CNOT basis gate as the lowest level of control in a quantum system---akin to NAND gates in digital logic.

However, the lowest level of quantum control is actually composed of analog pulses emitted by Arbitrary Waveform Generators (AWGs). Access to pulse-level control via frameworks like Qiskit Pulse \cite{mckay2018qiskit, alexander2020qiskit} enable programmers to act at the level of the quantum computer's \textit{Hamiltonian}, which describes possible energy configurations for the underlying qubits. Within this framework, CNOT is merely a pre-calibrated sequence of pulses that affect the Hamiltonian. The constituent pulses within the CNOT sequence are known as \textit{native gates} and correspond to the natural interactions on and between qubits, per the device Hamiltonian. Unlike the universal basis gates in OpenQASM, native gates are specific to the underlying hardware technology.

For example, superconducting qubits with tunable qubit energies (frequencies) typically implement native gates known as CPHASE and iSWAP \cite{krantz2019quantum}. Both Google \cite{arute2019quantum} and Rigetti \cite{smith2016practical} implement these two native gates. On trapped ion hardware, the XX \cite{maslov2017basic} or Mølmer-Sørensen \cite{sorensen1999quantum} interactions are native gates.

In this paper, we focus on the Cross-Resonance native gate, which is applicable to fixed-frequency superconducting qubits. We focus on Cross-Resonance, because it is the native gate on IBM's hardware which (a) is currently the most widely used quantum computing platform and (b) exposes access to native gates via Qiskit Pulse control.

\section{Warm-up: SWAP Orientation} \label{sec:swap_orientation}
We begin with a simple optimization that chooses the orientation of a SWAP to improve its speed and fidelity. This optimization relies on knowledge of the underlying native gates that make up the CNOT and NOTC basis gates.

As shown in Figure~\ref{fig:basic_swap}, a SWAP can be decomposed as either CNOT-NOTC-CNOT or as NOTC-CNOT-NOTC. We refer to the difference between CNOT versus NOTC as orientation. At the granularity of basis gates, the two orientations are equivalent. For example, OpenQASM v2 makes no distinction between CNOT and NOTC: they are treated as equals in terms of latency and fidelity. As such, from the programmer's view, there appears to be no difference between CNOT-NOTC-CNOT and NOTC-CNOT-NOTC.

However, a deeper understanding of the native gates underlying CNOT and NOTC reveals that hardware has a preferred orientation. Specifically, CNOTs on IBM's superconducting qubits are implemented with a sequence of three native Cross-Resonance (CR) gates. The CR native gate involves driving one of two connected qubits at the energy (frequency) of the other qubit. While CR could be implemented in either direction, in practice, it is faster and more reliable to apply the CR to the qubit with a higher underlying frequency \cite{rueschlikon, calibrating_two_qubit_gate}. In the next section, we will delve deeper into the CR native gate, but for now it suffices to understand that CR is a directed interaction.

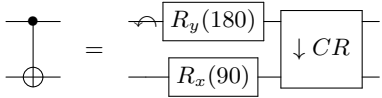
\begin{figure}[h]
$$
\Qcircuit @C=0.7em @R=0.7em {
& \Ctrl{1} & \qw &&\raisebox{-2.6em}{=}&&& {\curvearrowleft} \qw & \gate{R_y(180)} & \multigate{1}{\downarrow CR} & \qw \\
& \Targ    & \qw && &&& \qw & \gate{R_x(90)} &              \ghost{\downarrow CR}        & \qw
}
$$
    \caption{CNOT in the same orientation as the Cross Resonance (CR) native gate. We use curved/circled arrows to denote the virtual $R_z$ gates which are free. The $\downarrow$ emphasizes that CR is directed.} \label{fig:fast_cnot}
\end{figure}

Without loss of generality, we assume throughout this paper that the CR native gate between two qubits is directed to align with the CNOT orientation---control qubit on top and target qubit on the bottom. Figure~\ref{fig:fast_cnot} shows the resulting decomposition of the CNOT in terms of CR. In all circuit diagrams in this paper, we use the following notation for specific $R_z$ rotations:
$$\curvearrowleft := R_z(-90) \qquad \curvearrowright := R_z(90)  \qquad \circlearrowright := R_z(180)$$
We adopt this notation because $R_z$-type gates are implemented \textit{virtually} with perfect fidelity and zero latency on IBM's systems through classical bookkeeping \cite{mckay2017efficient}. The curved arrow notation is narrower than $\scriptsize \Qcircuit @C=0.3em @R=0.7em {& \gate{R_z(\theta)} & \qw}$, which visually emphasizes that $R_z$ gates are instantaneous, unlike $R_x$ and $R_y$ gates.

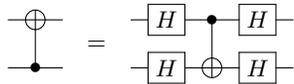
\begin{figure}[h]
$$
\Qcircuit @C=0.7em @R=0.7em {
& \Targ     & \qw &&\raisebox{-2.2em}{=}&&& \gate{H} & \Ctrl{1} & \gate{H} & \qw \\
& \Ctrl{-1} & \qw && &&& \gate{H} & \Targ    & \gate{H} & \qw
}
$$
    \caption{NOTC can be realized by sandwiching CNOT with Hadamard ($H$) single-qubit rotation gates.} \label{fig:flip_cnot}
\end{figure}

Since the CR is directed, we cannot achieve a NOTC by simply flipping the two wires in Figure~\ref{fig:fast_cnot}. Instead, we can rely on the well-known Hadamard sandwich identity \cite{nielsen2002quantum}. As depicted in Figure~\ref{fig:flip_cnot}, this identity allows us to execute NOTC by placing the CNOT circuit within four single-qubit rotation gates known as Hadamard's ($H$). The $H$ gate can be implemented by a combination of $R_x$ and $R_y$ gates, though the exact decomposition is not important here.

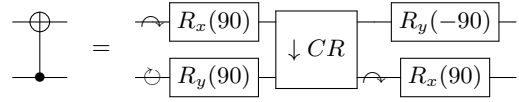
\begin{figure}[h]
$$
\Qcircuit @C=0.7em @R=0.7em {
& \Targ     & \qw &&\raisebox{-2.4em}{=}&&& \curvearrowright \qw & \gate{R_x(90)} & \multigate{1}{\downarrow CR} & \qw & \gate{R_y(-90)} & \qw \\
& \Ctrl{-1} & \qw && &&& \circlearrowright \qw & \gate{R_y(90)} & \ghost{\downarrow CR} & \curvearrowright \qw & \gate{R_x(90)} & \qw
}
$$
\caption{Decomposition of NOTC, given the directed CR.} \label{fig:slow_notc}
\end{figure}

Applying this identity to the CR decomposition of CNOT, we obtain the NOTC circuit in Figure~\ref{fig:slow_notc}. This is indeed how NOTC is implemented on IBM hardware.

Denoting the latency of single-qubit operations as $t_{1q}$ and the latency of Cross-Resonance as $t_{CR}$, we see that the CNOT in Figure~\ref{fig:fast_cnot} has a runtime of $t_\text{CNOT} = t_{1q} + t_{CR}$. By contrast, the NOTC in Figure~\ref{fig:slow_notc} has a runtime of $t_\text{NOTC} = 2t_{1q} + t_{CR}$. Therefore, it is always faster to SWAP with the CNOT-NOTC-CNOT orientation than with the NOTC-CNOT-NOTC orientation. From the perspective of traditional quantum programming, this is a surprising result since SWAP is a fundamentally symmetric operation. It is only by understanding the hardware-level native gate primitives that we can see the orientation asymmetry between CNOT and NOTC.

The speedup factor achieved from picking the faster orientation is

\begin{equation}
\frac{t_{\text{NOTC-CNOT-NOTC}}}{t_{\text{CNOT-NOTC-CNOT}}} = \frac{t_\text{CNOT} + 2 t_\text{NOTC}}{2t_\text{CNOT} + t_\text{NOTC}} \frac{}{} = \frac{5t_{1q} + 3t_{CR}}{4t_{1q} + 3t_{CR}}
    \label{eq:speedup}
\end{equation}

The exact speedup depends on the relative ratio of $t_{CR}$ and $t_{1q}$, which in turn is largely dependent on the coupling strength between pairs of connected qubits. On IBMQ's Johannesburg (20 qubits, 23 connected pairs) and Paris (27 qubits, 28 connected pairs) devices, the mean speedups are 2.5\% and 3.2\% respectively. It bears emphasizing that due to the short \textit{coherence lifetimes} of qubits, speeding up a quantum computation translates directly into improving the probability of success. Since qubit quality decays exponentially with the computation duration, speedups are particularly critical to boost the success of quantum programs.

Aside from the speedup gains, the CNOT-NOTC-CNOT orientation is also preferable because it requires fewer single qubit rotations. To quantify this, we count the \textbf{active rotation} over the single-qubits gates in a circuit. We include $R_x$ and $R_y$ gates in the total, but exclude $R_z$ gates which are implemented virtually with perfect fidelity. Looking at Figure~\ref{fig:fast_cnot}, we see 180 + 90 = 270 degrees of active single-qubit rotation. However, as explained later in Section~\ref{sec:optimized_swap}, the $\downarrow CR$ block internally performs an additional 180 degrees of active single-qubit rotation. Thus, CNOT requires 270 + 180 = 450 degrees of active rotation. By contrast, the NOTC in Figure~\ref{fig:slow_notc} requires 360 + 180 = 540 degrees of active rotation. Therefore, the fast CNOT-NOTC-CNOT orientation has only 1440 degrees of active single-qubit rotation, whereas the slow NOTC-CNOT-NOTC orientation requires 1530 degrees. This 90 degree saving is small but meaningful, given how frequent SWAP gates are.

While the SWAP Orientation optimization is motivated and informed by observations regarding native gates, it can be implemented without actual programmer access to native gates (e.g. via Qiskit Pulse). As long as programmers are aware of the direction of the underlying CR native gates, they can correctly orient all SWAPs to occur in the CNOT-NOTC-CNOT orientation. This orientation both achieves a speedup (per Equation~\ref{eq:speedup}) and saves 90 degrees unnecessary active single-qubit rotation.


\section{Optimized SWAP} \label{sec:optimized_swap}
We now extend the warm-up in Section~\ref{sec:swap_orientation} by delving further into the native gates. Unlike the SWAP Orientation optimization which can be implemented with standard basis gates, all of the optimizations here require programmer access to native gates---e.g. through Qiskit Pulse. Table~\ref{tab:SWAP} summarizes our optimizations in terms of depth (i.e. runtime, in terms of the single-qubit gate duration $t_{1q}$ and CR native gate duration $t_\text{CR}$) and degrees of active single-qubit rotation. The depths and active rotations for Slow Orientation (NOTC-CNOT-NOTC) and Fast Orientation (CNOT-NOTC-CNOT) are as presented in Section~\ref{sec:swap_orientation}. 

\begin{table}[h]
\small
\renewcommand{\arraystretch}{1.4} \centering
\begin{tabular}{P{0.17\textwidth}|p{0.12\textwidth}|p{0.13\textwidth}}
Technique         & Depth & Active 1-Qubit Rotation          \\ \hline
\textbf{Slow Orientation} & $\mathbf{5t_{1q} + 3t_{CR}}$ & \textbf{990 + 540} \\
Fast Orientation & $4t_{1q} + 3t_{CR}$ & 900 + 540\\
Cross-Gate Pulse Cancellation & $3t_{1q} + 3t_{CR}$ & 720 + 540 \\
Commutation through CR & $3t_{1q} + 3t_{CR}$ & 450 + 540 \\
\textbf{CR Polarity} & $\mathbf{2t_{1q} + 3t_{CR}}$ & \textbf{270 + 540}
\end{tabular}
\caption{Optimizations for SWAP. Our experimental evaluation compares the first row, Slow Orientation, to the last row, CR Polarity. We refer to these as Standard and Optimized SWAP respectively.}
\label{tab:SWAP}
\end{table}

The last three rows correspond to the three native gate based optimizations presented in this section: Cross-Gate Pulse Cancellation, Commutation through CR, and CR Polarity. Our experimental results in Section~\ref{sec:results} compare the Slow Orientation row versus the CR Polarity row. Henceforth, we refer to these as Standard SWAP and Optimized SWAP respectively.

\begin{figure}[h]
    \centering
    \includegraphics[width=0.5\textwidth]{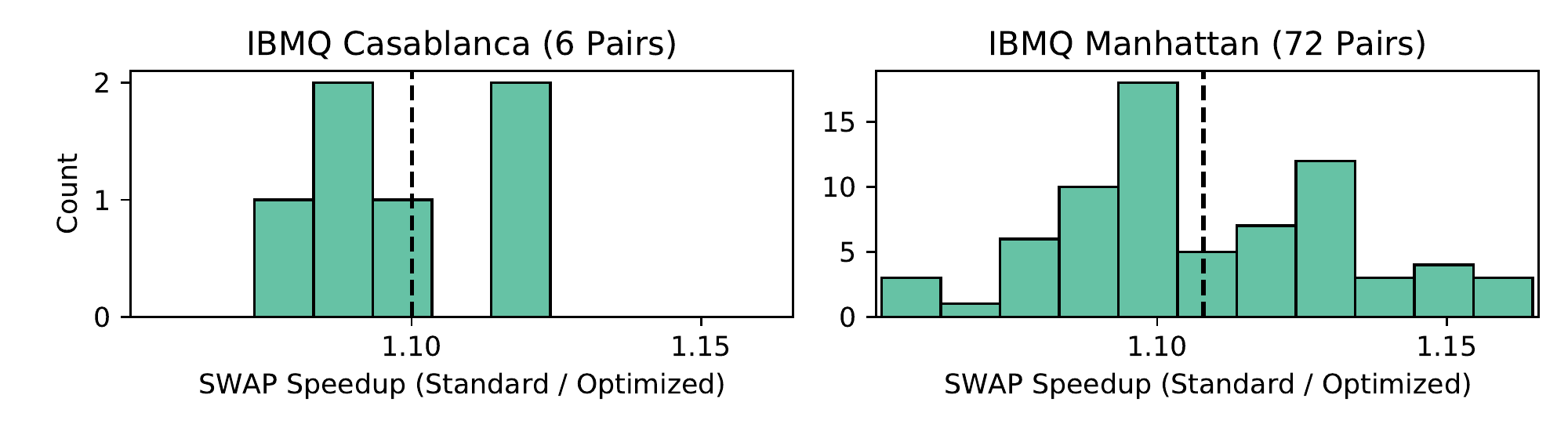}
    \caption{Histograms of SWAP speedups via our Optimized SWAP versus the Standard SWAP. Data from 7-qubit Casablanca device (with 6 connected pairs) and from 65-qubit Paris device (with 72 connected pairs). The mean speedup factors (dashed vertical lines) are 1.10 and 1.11 respectively.}
    \label{fig:speedup_histogram}
\end{figure}

Per Table~\ref{tab:SWAP}, the speedup factor of the Optimized SWAP relative to the Standard SWAP is $(5t_{1q} + 3t_{CR}) / (2t_{1q} + 3t_{CR})$. The exact speedup varies across each pair of connected qubits. Figure~\ref{fig:speedup_histogram} presents a histogram of the exact speedups across the 6 connected qubit pairs on IBMQ Casablanca and the 72 connected pairs on IBMQ Manhattan. The mean speedups are 10\% and 11\% respectively. In addition to the speedup, the Optimized SWAP only requires 270 + 540 degrees of active single-qubit rotation versus 990 + 540 degrees for the Standard SWAP. The 720 degrees of rotation savings correspond to unnecessary single qubit rotation gates and therefore, avoidance of error accumulation.


\subsection{Cross-Gate Pulse Cancellation} \label{subsec:cgpc}

\begin{figure}[h]
$$ \footnotesize
\Qcircuit @C=0.6em @R=1.7em {
& \color{blue} \curvearrowleft \qw & \gate{\color{blue} R_y(180)} & \multigate{1}{\color{blue} \downarrow CR} & \color{red} \curvearrowright \qw & \gate{\color{red} R_x(90)} & \multigate{1}{\color{red} \downarrow CR} & \qw & \gate{\color{red} R_y(-90)} & \color{blue} \curvearrowleft \qw & \gate{\color{blue} R_y(180)} & \multigate{1}{\color{blue} \downarrow CR} \\
& \qw & \gate{\color{blue} R_x(90)} & \ghost{\color{blue} \downarrow CR} & \color{red} \circlearrowright \qw & \gate{\color{red} R_y(90)} & \ghost{\color{red} \downarrow CR} & \color{red} \curvearrowright \qw & \gate{\color{red} R_x(90)} & \qw & \gate{\color{blue} R_x(90)} & \ghost{\color{blue} \downarrow CR} \gategroup{1}{9}{1}{11}{0.7em}{--} \gategroup{2}{9}{2}{11}{0.7em}{--}
}
$$
    \caption{Fast orientation of SWAP expressed in terms of the pulse-level $\downarrow CR$ primitives. Cross-gate pulse cancellation allows simplification of the rotation gates in the dashed boxes.} \label{fig:fsf_swap}
\end{figure}
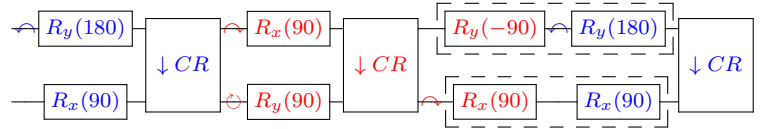

We take the Fast Orientation (CNOT-NOTC-CNOT) as the starting point for our optimizations. The first native gate based optimization we invoke is called Cross-Gate Pulse Cancellation, as introduced by Section 5 of our previous work \cite{gokhale2020optimized}. The core motivation behind this technique is that by treating CNOT (NOTC) as a basis gate, no further optimizations are possible to the CNOT-NOTC-CNOT sequence. However, when we decompose down to the CR native gates, new opportunities for gate cancellation emerge. To this end, we begin by concatenating the gate sequences for CNOT, NOTC, and CNOT, using the decompositions in Figure~\ref{fig:fast_cnot}~and~\ref{fig:slow_notc}. Figure~\ref{fig:fsf_swap} shows the result, with the CNOTs in blue and the NOTC in red to indicate the boundaries.

As indicated by the two dashed boxes in Figure~\ref{fig:fsf_swap}, this concatenated view reveals new opportunities for optimization that are invisible when we consider the CNOT and NOTC as atomic basis gates. This observation is akin to interprocedural optimization in classical compilers \cite{cooper1986interprocedural, hall1991managing} and to fine-grained scheduling in quantum compilers \cite{heckey2015compiler}. On the top qubit, the $\scriptsize \Qcircuit @C=0.7em @R=0.5em {& \gate{R_y(-90)} & \curvearrowleft \qw & \gate{R_y(180)} \qw }$ sequence can be simplified into a single $\scriptsize \Qcircuit @C=0.7em @R=0.5em {& \curvearrowright \qw & \gate{R_x(-90)} \qw }$ sequence. This optimization means that the qubit only actively rotates through 90 degrees instead of 270 degrees (recall that the $R_z$ gates are virtual and do not incur any active rotation). On the bottom qubit, the $\scriptsize \Qcircuit @C=0.7em @R=0.5em {& \gate{R_x(90)} & \gate{R_x(90)} & \qw }$ sequence can be simplified into a single $\scriptsize \Qcircuit @C=0.7em @R=0.5em {& \gate{R_x(180)} & \qw }$ gate. Although it does not reduce the total amount of rotation on the bottom qubit, the single-pulse sequence is faster, less susceptible to calibration error, and achieves higher fidelity experimentally as shown in \cite{gokhale2020optimized}.

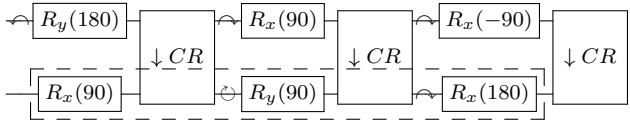
\begin{figure}[h]
$$ \footnotesize
\Qcircuit @C=0.6em @R=1.7em {
& \curvearrowleft \qw & \gate{R_y(180)} & \multigate{1}{\downarrow CR} & \curvearrowright \qw & \gate{R_x(90)} & \multigate{1}{\downarrow CR} & \curvearrowright \qw & \gate{R_x(-90)} & \multigate{1}{\downarrow CR} \\
& \qw & \gate{R_x(90)} & \ghost{\downarrow CR} & \circlearrowright \qw & \gate{R_y(90)} & \ghost{\downarrow CR} & \curvearrowright \qw & \gate{R_x(180)} & \ghost{\downarrow CR} \gategroup{2}{3}{2}{9}{.7em}{--}
}
$$
    \caption{SWAP after cross-gate pulse cancellation. The gates in the dashed box on the bottom qubit will be optimized next.} \label{fig:post_cgpc}
\end{figure}

Figure~\ref{fig:post_cgpc} shows the resulting pulse sequence for the SWAP after applying these pulse gate cancellations. In addition to the fact that the top qubit avoids 180 degrees of unnecessary rotation and the single-pulse improvement on the bottom qubit, the entire sequence is shallower than Figure~\ref{fig:fsf_swap} in depth by $t_{1q}$. No further improvement is possible until we stop treating the $\downarrow CR$ native gate as a black box. This motivates the next optimization, Commutation through CR.

\subsection{Commutation through CR} \label{subsec:cr_commutativity}

\begin{figure}[h]
$$ \scriptsize
\Qcircuit @C=0.5em @R=1em {
& \multigate{1}{\downarrow CR} & \qw & \raisebox{-2.6em}{=} && \ctrlo{1}      & \Ctrl{1}        & \gate{R_x(180)} & \Ctrl{1} &  \ctrlo{1} & \qw \gategroup{1}{6}{2}{7}{.7em}{--} \gategroup{1}{9}{2}{10}{.7em}{--}  \\
& \ghost{\downarrow CR}        & \qw &   && \gate{R_x(45)} & \gate{R_x(-45)} & \qw             & \gate{R_x(45)} & \gate{R_x(-45)} & \qw \\
& &  &&& \mbox{\quad \qquad \qquad positive half-CR}      & & & \mbox{\quad \qquad \qquad negative half-CR} & &  
}
$$
\caption{The CR native gate is implemented with an echoed sequence that applies a positive and then negative half-CR, with $R_x(180)$ in between. This echo results in cancellation of unwanted components. Purely as a conceptual aid, we decompose the half-CRs into intuitive gate implementations emphasizing that the bottom qubit is only acted upon by $R_x$ gates.}
\label{fig:cross_resonance}
\end{figure}
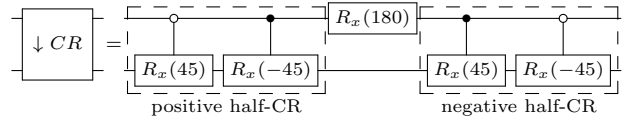

To optimize further, we must examine the operation of the CR native gate more closely. Figure~\ref{fig:cross_resonance} depicts a circuit-level view of CR between two qubits. Notice that CR is implemented in an echoed fashion that first applies a positive half-CR pulse and then a negative half-CR pulse. In between the two is a $R_x(180)$ gate on the top qubit \footnote{This $R_x(180)$ gate is what accounts for the additional $540 = 3 \times 180$ degrees of active rotation in Table~\ref{tab:SWAP}}, which effectively turns the negative half-CR pulse into a positive one. This results in a positive full-CR pulse, but with protection against some coherent errors. This is because unwanted components in the negative CR native gate maintain their negative sign and cancel out with the positive CR native gate. We refer to \cite{krantz2019quantum, sheldon2016procedure} for further details about echoed Cross Resonance.

It bears mentioning that technically, Figure~\ref{fig:cross_resonance} implements CR \textit{plus} a side effect of $R_x(180)$ on the top qubit. This is due to the sandwiched $R_x(180)$ needed for echo cancellation. However, in creating a CNOT or NOTC, the side effect is handled by the single qubit rotations in Figures~\ref{fig:fast_cnot} and ~\ref{fig:slow_notc}.

The circuitry inside the dashed boxes is purely a conceptual aid. In reality, the dashed boxes are realized by a more fundamental Hamiltonian interaction that is not captured by the gate model. However, this gate level view is functionally equivalent. To perform the next optimization, we note that the bottom qubit of the Cross-Resonance interaction is only affected by $R_x$ gates. Therefore, $R_x$ gates on one side of a CR's target can be moved to the opposite side, because consecutive $R_x$ gates can be freely interchanged with each other. Formally, this is a commutativity relationship.

\begin{figure}[h]
$$ \footnotesize
\Qcircuit @C=0.6em @R=1.7em {
& \curvearrowleft \qw & \gate{R_y(180)} & \multigate{1}{\downarrow CR} & \curvearrowright \qw & \gate{R_x(90)} & \multigate{1}{\downarrow CR} & \curvearrowright \qw & \gate{R_x(-90)} & \multigate{1}{\downarrow CR} \\
& \qw & \qw & \ghost{\downarrow CR} & \curvearrowleft \qw & \gate{R_x(90)} & \ghost{\downarrow CR} & \curvearrowleft \qw & \qw & \ghost{\downarrow CR} \gategroup{1}{3}{2}{6}{0.7em}{--}
}
$$
    \caption{After commuting $R_x$ rotations in the dashed box in Figure~\ref{fig:post_cgpc} to the zone between the first two $\downarrow CR$'s, cancellation leads to $R_z(-90)$ followed by $R_x(90)$. The gates in the left box will be optimized next.}
    \label{fig:post_commutation}
\end{figure}
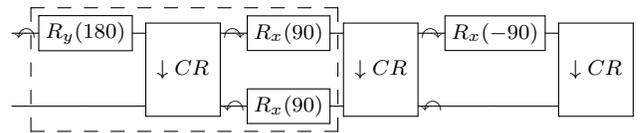

Consequently, the $R_x$ gates inside the left and right\footnote{The right $R_x(180)$ has a $R_z(90)$ in the way, but can be commuted using the identity that $R_x(180)R_z(90) = R_z(-90)R_x(180)$.} sides of the dashed box in Figure~\ref{fig:post_cgpc} can be moved to the center, in between the first two $\downarrow CR$'s. After the commutation, the gates between the first two $\downarrow CR$'s are, from left to right: $\scriptsize \Qcircuit @C=0.7em @R=0.5em {& \gate{R_x(90)} & \circlearrowright \qw & \gate{R_y(90)} & \gate{R_x(180)} & \qw }$. It can be shown that this gate sequence compresses to $\scriptsize \Qcircuit @C=0.7em @R=0.5em {& \curvearrowleft \qw & \gate{R_x(90)} & \qw }$, which only requires 90 degrees of active rotation, as opposed to 360 degrees. The resulting pulse sequence is shown in Figure~\ref{fig:post_commutation}. Note that there are no savings in depth (runtime) relative to Figure~\ref{fig:post_cgpc}.

\subsection{Cross-Resonance Polarity} \label{subsec:polarity}

Our final optimization begins with the observation that the echoed CR native gate does not necessarily need to be ordered [positive half-CR, negative half-CR]. We can instead implement the echoed CR native gate as [negative half-CR, positive half-CR]. We refer to this as a polarity switch. A polarity switch does however create a new side effect---specifically enacting a $R_x(180)$ gate to the left of the CR native gate. However, this extra side effect can be beneficial and lead to cancellation. In particular, notice that the leftmost CR in the dashed box of Figure~\ref{fig:post_commutation} has a $R_y(180)$ gate on the top left. Switching the polarity of this CR native gate causes beneficial cancellation between the new side effect and this $R_y(180)$ gate.

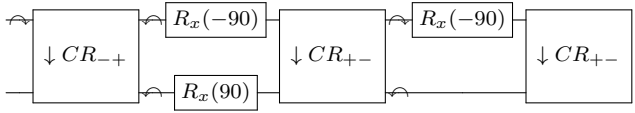
\begin{figure}[h]
$$ \footnotesize
\Qcircuit @C=0.6em @R=1.7em {
& \curvearrowright \qw & \multigate{1}{\downarrow CR_{-+}} & \curvearrowleft \qw & \gate{R_x(-90)} & \multigate{1}{\downarrow CR_{+-}} & \curvearrowright \qw & \gate{R_x(-90)} & \multigate{1}{\downarrow CR_{+-}} \\
& \qw & \ghost{\downarrow CR_{-+}} & \curvearrowleft \qw & \gate{R_x(90)} & \ghost{\downarrow CR_{+-}} & \curvearrowleft \qw & \qw & \ghost{\downarrow CR_{+-}}
}
$$
    \caption{Final Optimized SWAP, after CR polarity switch on the first CR native gate. The $-+$ subscript denotes that the first CR has switched polarity, whereas the other two maintain $+-$ polarity. For actual execution, we also converted the $R_x(-90)$ gates into $R_x(90)$ gates via virtual $R_z$ rotations; this is shown in Figure~\ref{fig:optimized_swap_with_x90} in the Appendix.}
    \label{fig:optimized_swap}
\end{figure}

Figure~\ref{fig:optimized_swap} shows the circuit after applying a polarity switch on the first CR native gate. This is the final Optimized SWAP. Changing the polarity of the other two CR native gates does not improve the circuit. Examining the final circuit, we see that it has a depth of $2t_{1q} + 3t_{CR}$ and only 270 degrees of active single qubit rotation (plus the $540 = 3 \times 180$ degrees inside of the CR's). To the best of our knowledge, no further zero-calibration optimizations are possible. The Optimized SWAP is available through the SuperstaQ platform \cite{superstaq}.

\section{Experimental Results on IBM Hardware} \label{sec:results}
\subsection{Evaluation of Optimized SWAP Gate}
\begin{figure}[h]
    \centering
    \includegraphics[width=0.49\textwidth]{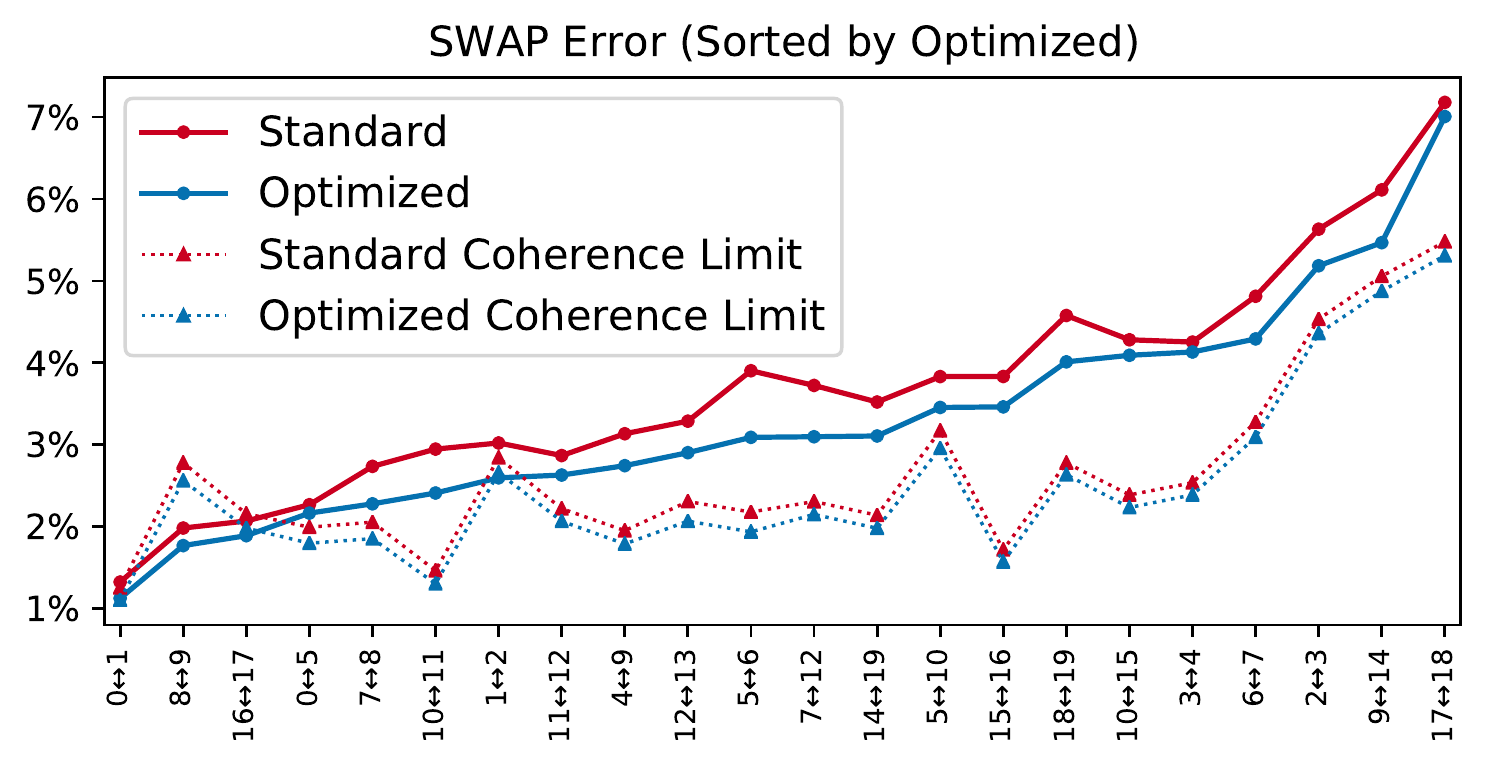}
    \caption{Error of Standard (red) and Optimized (blue) SWAPs across 22 pairs of connected qubits on IBMQ Johannesburg, calculated via Interleaved Randomized Benchmarking. The pairs are sorted by the Optimized SWAP error. As expected, the Optimized SWAP always has lower error than the Standard SWAP. The dashed lines correspond to the minimum error for each pair, via the coherence-limited gate error calculation.}
    \label{fig:swap_error}
\end{figure}

\begin{figure}[h]
    \centering
    \includegraphics[width=0.49\textwidth]{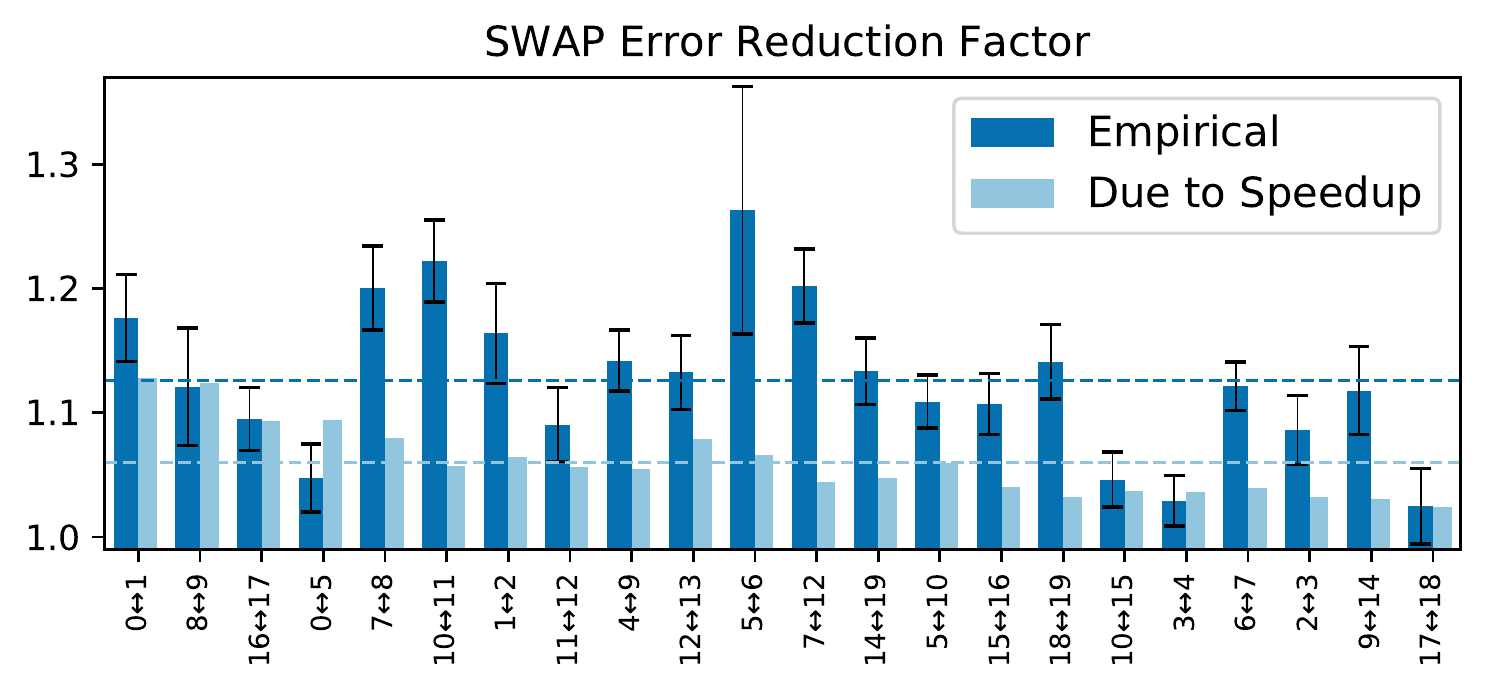}
    \caption{SWAP error reduction factor, showing empirical data as well as the fraction of the empirical speedup that is attributed to the Optimized gate being faster.}
    \label{fig:swap_error_reduction}
\end{figure}

We used Interleaved Randomized Benchmarking (IRB) \cite{magesan2012efficient} to measure the fidelity of Optimized and Standard SWAPs on each pair of connected qubits. Raw IRB results are provided in Appendix~\ref{app:rb}. Here in the main text, we begin with the SWAP errors calculated from IRB. Figure~\ref{fig:swap_error} plots these errors for Standard and Optimized SWAPs on each of the 22 qubit pairs on the IBMQ Johannesburg device. The pairs are sorted by Optimized SWAP error. The mean error of the Optimized SWAP is 3.3\%, versus 3.7\% for the Standard SWAP. This is an average $\sim$ 13\% reduction in SWAP error, which experimentally validates our optimizations. Moreover, as we experimentally demonstrate in Section~\ref{sec:results}, the 13\%  reduction in per-SWAP error can compound into much larger reductions on benchmarks that require several SWAPs. Note that we omit error bars in Figure~\ref{fig:swap_error} for readability, but we will consider uncertainty in Figure~\ref{fig:swap_error_reduction}.

Figure~\ref{fig:swap_error} also has dashed lines plotting the \textit{coherence-limited gate error} calculation \cite{gambetta2012characterization, Qiskit} for each qubit pair. This calculation lower bounds the gate error based on the fact that qubits have finite coherence time, and therefore qubits decohere during the gate. The calculation, plotted with a dashed line, is based on the gate duration of SWAPs on each pair as well as the coherence times ($T_1$ and $T_2$) of each qubit.

In general, the errors based on coherence limits (dashed lines) account for about half of the the empirically observed SWAP errors (solid lines). This agrees with IBM's latest hardware evaluations \cite{gambetta19cramming, sundaresan2020reducing}. Once gates achieve the error lower bounds set by coherence limits, subsequent progress will require either faster gates (e.g. by techniques like our Optimized SWAP or by driving AWGs at higher amplitudes) or longer qubit coherence times.

We conclude this section by estimating and understanding the exact error reduction factor of the Optimized SWAP. The bar plot in Figure~\ref{fig:swap_error_reduction} shows the empirical ratio between the SWAP Errors of Standard and Optimized for each pair. We include $\pm \sigma$ error bars (omitted in the previous plots for readability) based on the errors of the IRB exponential decay fit in Appendix~\ref{app:rb} and the calculation for the variance of a ratio \cite{seltman2012approximations}. Figure~\ref{fig:swap_error_reduction} also includes bars capturing how much of the error reductions are attributable to the speedup of the Optimized SWAP relative to the Standard SWAP. These bars are based on the coherence-limited gate error calculations and correspond to the gap between the red and blue dashed lines in Figure~\ref{fig:swap_error}.

On all but one \footnote{The exception is the $17 \leftrightarrow 18$ pair. The point estimate for the SWAP error reduction is 1.02 which is `only' $0.8\sigma$ better than 1.0 parity. We attribute this to the fact that $17 \leftrightarrow 18$ pair is also the pair with highest absolute error.} of the qubit pairs, the Optimized SWAP is better than the Standard SWAP by at least a $1\sigma$ margin. On average, the Optimized SWAP reduces error by a factor of 1.13. The speedup (lower depth) of the Optimized SWAP accounts for an error reduction factor of 1.06---about half of the total empirical gain. The remaining half should be attributed to the fact that the Optimized SWAP requires only 270 + 540 degrees of active single-qubit rotation, versus 990 + 540 degrees for the Standard SWAP. Notice that on three pairs of qubits ($8 \leftrightarrow 9$, $0 \leftrightarrow 5$, $3 \leftrightarrow 4$), the theoretical error reduction due to the speedup exceeds the empirically realized reduction. This discrepancy can likely be explained by the error bars on the empirical error reduction.

\subsection{Application-Level Benchmarks} \label{subsec:application_level_benchmarks}
We experimentally evaluated our Optimized SWAP on a suite of application-level benchmarks. For each benchmark, we ran 8000 shots with our native gate based optimizations and 8000 shots with standard basis gate based approaches. We invoked the maximum \texttt{optimization\_level} in Qiskit's circuit transpiler to map logical to physical qubits in way that (a) minimizes qubit communication cost based on hardware connectivity and (b) prefers higher quality qubits. Note that (a) ensures that we compare to a fair baseline that does not have artificially high qubit communication cost. We also set a transpiler randomization seed to ensure reproducibility.

Our optimizations were tested on four benchmark types: (1) Bernstein-Vazirani which aims to detect a `secret' all-ones bitstring \cite{bernstein1997quantum}, (2) Generalized Toffoli \cite{gokhale2019asymptotic, gokhale2020extending} which computes the AND of input qubits, (3) quantum Adder \cite{qiskit2020weighted} which computes $\ket{11} + \ket{11} = \ket{110}$, and (4) Long SWAP which chains SWAPs to move a source qubit to a distant target, i.e. $\ket{100...000} \rightarrow \ket{000...001}$. To avoid overloading the queue on any specific machine, we executed Bernstein-Vazirani on IBMQ Johannesburg (20 qubit device), Generalized Toffoli and Long SWAP on IBMQ Paris (27 qubit device), Adder on IBM Q Bogota (5 qubit device).

\begin{figure}[h]
    \centering
    \includegraphics[width=0.49\textwidth]{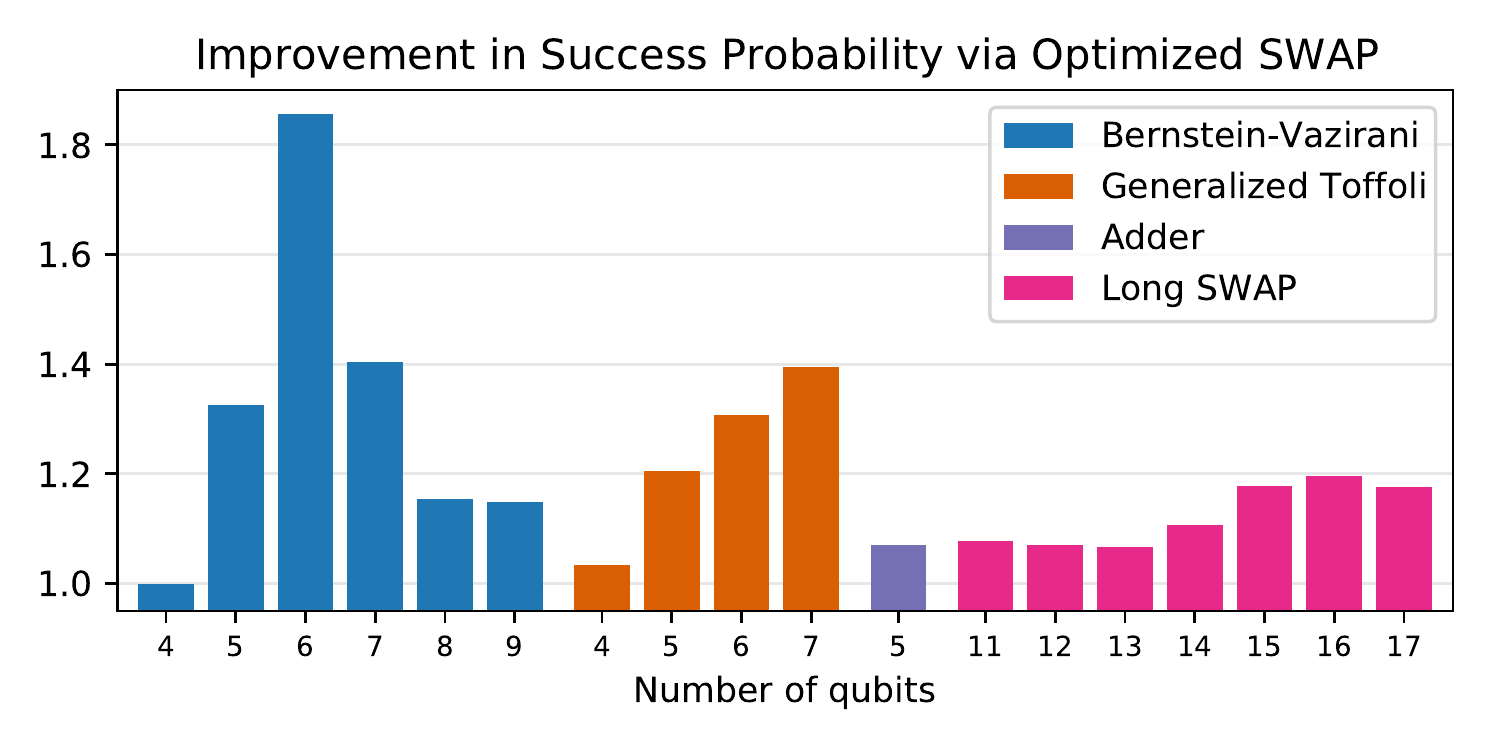}
    \caption{Benchmark results across four applications. The Optimized SWAP boosts typical success by 10--40\%.}
    \label{fig:PST_Benchmarks}
\end{figure}

\begin{figure}
    \centering
    \includegraphics[width=0.49\textwidth]{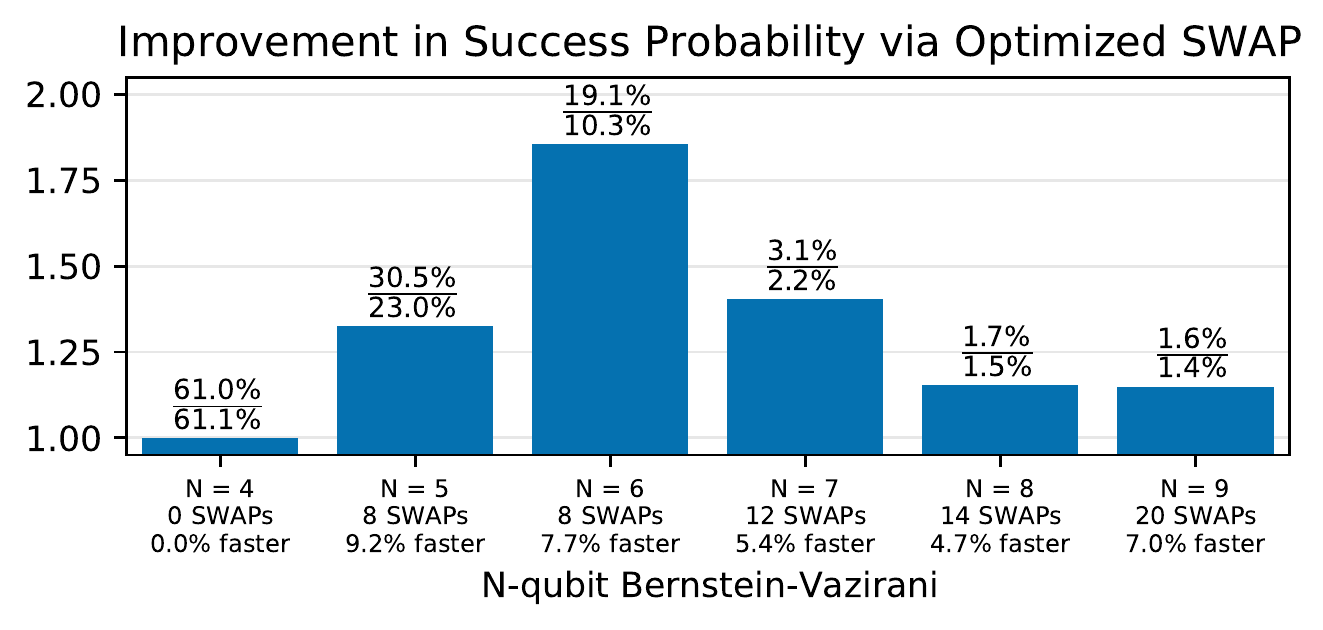}
    \caption{Zoom-in of Bernstein-Vazirani improvement in success (measuring all-ones secret) with $N = 4$ through $9$ qubits. The fractions show \% success via Optimized SWAP divided by \% success via Standard SWAP. Adding more SWAPs improves our gain, until other noise effects dominate.}
    \label{fig:BV}
\end{figure}

Figure~\ref{fig:PST_Benchmarks} shows benchmark results. The Optimized SWAP boosts typical success rates by 10--40\%, with a maximum boost of 85\% for 6-qubit Bernstein-Vazirani. To better understand the results, we zoom in on the Bernstein-Vazirani sweep from $N = 4$ to 9 qubits in Figure~\ref{fig:BV}. The Figure includes exact Optimized and Standard success probabilities, the number of SWAPs, and the total circuit speedup due to optimization. The $N = 4$ case is a control trial that has 0 SWAPs: as expected, Standard and Optimized are identical. As we go to $N = 5$ and 6, the improvement factor rises, before falling from $N = 7$ to 9. We attribute this pattern to the following. As we increase the benchmark size, the number of SWAPs increases, leading to more multiplicative compounding gains in the relative success between Optimized and Standard. However, once we have benchmarks that are too large, the circuit output is dominated by other noises (readout noise, non-SWAP gate errors, etc.) and the advantage cannot be discerned. As hardware noise continues to diminish, we expect that Optimized SWAPs will perform even better for larger benchmark sizes.

\subsection{Analysis} \label{subsec:analysis}

When we use our Optimized SWAP on full benchmarks, we expect improved experimental results for two reason. First, the Optimized SWAP has mean fidelity of 0.967 versus 0.963 for the Standard SWAP. To good approximation, gate errors accumulate multiplicatively, so we therefore expect a $\sim (0.967 / 0.963) ^ k$ factor improvement in a circuit with $k$ SWAPs. However, this alone does not nearly account for the gains seen in our benchmarks. For example, $k = 10$ SWAPs would only yield a 4\% improvement. This brings us to the second reason for improved experimental results: the Optimized SWAP leads to shorter total circuit duration. This means less time for each qubit to decohere---even on idle qubits that are not involved in SWAPs. For a circuit of duration $T$, the errors accumulated on \textit{each} qubit due to decoherence are $1 - e^{-T/T_1}$ and $1 - e^{-T/T_2}$ for $T_1$ and $T_2$ qubit lifetimes respectively. Putting these together, we estimate a total improvement factor of:
\begin{equation}
 \left(\frac{1-\text{error}_\text{Opt}}{1 - \text{error}_\text{Std}}\right)^k \times \left( e^{\Delta T / T_1 + \Delta T / T_2} \right)^N
 \label{eq:model}
\end{equation}
for a $N$-qubit circuit with $k$ SWAPs and a total runtime speedup of $\Delta T = T_\text{Std} - T_\text{Opt}$. Note the last term is exponential in $N$.

This model roughly captures the behavior of our application-level benchmark results. For example, 
the Bernstein-Vazirani $N = 8$ benchmark had $k=14$ SWAPs and $\Delta T = 4.7\% \times 13.8$ \si{\micro}s = 0.63 \si{\micro}s, and the qubits on IBM Q Johannesburg have mean $T_1$ and $T_2$ lifetimes of about 75 \si{\micro}s. Plugging into Equation~\ref{eq:model} gives a success improvement factor of $\sim 1.21$ which is similar to the 1.15 factor observed experimentally. The $N = 6$ Bernstein-Vazirani appears to be an outlier that performs better experimentally than theory would predict. We attribute this to: (a) the Optimized SWAP error reduction factor in Figure~\ref{fig:swap_error_reduction} is generally higher on qubits with high connectivity connectivity qubits, which the qubit mapper prefers; and (b) day-to-day experimental variation on the device.

\section{Conclusion} \label{sec:conclusion}
We developed an Optimized SWAP that performing manipulating native gates. Our Optimized SWAP is 11\% faster and 13\% more reliable than the Standard SWAP. These speedups and error reductions will be universal across IBM devices, based on Figure~\ref{fig:speedup_histogram} and the analysis in Section~\ref{subsec:analysis}. In fact, our demonstration of the Optimized SWAP on the IBMQ Casablanca device resulted in the top-ranking submission to the IBM Quantum Open Science Challenge for better SWAP gates.

Recent research in Figure 3 of \cite{zlokapa2020boundaries} extrapolates hardware progress in superconducting qubits to forecast 23\% annual fidelity gains for two-qubit gates in superconducting hardware. Given that SWAPs account for almost all two-qubit gates in sparse hardware, we suggest that our Optimized SWAP in Qiskit Pulse could be compared to six months worth of hardware progress.

Moreover, the advantage of faster and reliable SWAPs compounds multiplicatively---both in the number of SWAPs due to the first term in Equation~\ref{eq:model} and in the number of qubits due to the second term. As such, we experience a supercharged Amdahl's Law for application-level benchmarks since a) SWAPs dominate two-qubit gates for typical programs and (b) improvements in gate fidelities compound multiplicatively. For example, the experimental results in Figure~\ref{fig:PST_Benchmarks} show that the Optimized SWAP boosts typical program success rates by 10--40\%.

Our Optimized SWAP is bootstrapped from pre-calibrated gates. This is important for three reasons. First, it means that our technique has zero calibration overhead and can be executed without requiring new calibration cycles or hampering system availability. As such, the Optimized SWAP has been deployed through the SuperstaQ platform \cite{superstaq}. Second, our technique has lower susceptibility to drift than what complex or high-frequency custom pulses would exhibit, e.g. due to varying transfer functions from room temperature control electronics to cold qubits \cite{alexander2020qiskit}. Finally, since our Optimized SWAP is bootstrapped from the pre-calibrated pulses, it can be encoded into a small payload with a simple encoding. By contrast, a complex custom SWAP pulse would need to be specified as a long time series, which could limit payloads for long pulse schedules.

\section*{Acknowledgements}
This work is supported by the U.S. Department of Energy, Office of Science, Office of Advanced Scientific Computing Research under Award Number DE-SC0021526 and the Accelerated Research in Quantum Computing (ARQC) program.

We acknowledge the use of IBM Quantum services for this work. The views expressed are those of the authors, and do not reflect the official policy or position of IBM or the IBM Quantum team.

We thank Ali Javadi-Abhari, Nate Earnest-Noble, and Victory Omole for helpful discussions.

Disclosure:  Fred Chong is Chief Scientist at Super.tech and an advisor to Quantum Circuits, Inc.

\bibliographystyle{unsrt}
\bibliography{references}

\begin{thebibliography}{10}

\bibitem{javadi2017optimized}
Ali Javadi-Abhari, Pranav Gokhale, Adam Holmes, Diana Franklin, Kenneth~R
  Brown, Margaret Martonosi, and Frederic~T Chong.
\newblock Optimized surface code communication in superconducting quantum
  computers.
\newblock In {\em Proceedings of the 50th Annual IEEE/ACM International
  Symposium on Microarchitecture}, pages 692--705, 2017.

\bibitem{siraichi2018qubit}
Marcos~Yukio Siraichi, Vin{\'\i}cius Fernandes~dos Santos, Sylvain Collange,
  and Fernando Magno~Quint{\~a}o Pereira.
\newblock Qubit allocation.
\newblock In {\em Proceedings of the 2018 International Symposium on Code
  Generation and Optimization}, pages 113--125, 2018.

\bibitem{zulehner2018efficient}
Alwin Zulehner, Alexandru Paler, and Robert Wille.
\newblock An efficient methodology for mapping quantum circuits to the {IBM QX}
  architectures.
\newblock {\em IEEE Transactions on Computer-Aided Design of Integrated
  Circuits and Systems}, 38(7):1226--1236, 2018.

\bibitem{tannu2019not}
Swamit~S Tannu and Moinuddin~K Qureshi.
\newblock Not all qubits are created equal: a case for variability-aware
  policies for {NISQ}-era quantum computers.
\newblock In {\em Proceedings of the Twenty-Fourth International Conference on
  Architectural Support for Programming Languages and Operating Systems}, pages
  987--999, 2019.

\bibitem{li2019tackling}
Gushu Li, Yufei Ding, and Yuan Xie.
\newblock Tackling the qubit mapping problem for {NISQ}-era quantum devices.
\newblock In {\em Proceedings of the Twenty-Fourth International Conference on
  Architectural Support for Programming Languages and Operating Systems}, pages
  1001--1014, 2019.

\bibitem{wille2019mapping}
Robert Wille, Lukas Burgholzer, and Alwin Zulehner.
\newblock Mapping quantum circuits to {IBM QX} architectures using the minimal
  number of {SWAP} and {H} operations.
\newblock In {\em 2019 56th ACM/IEEE Design Automation Conference (DAC)}, pages
  1--6. IEEE, 2019.

\bibitem{cowtan2019qubit}
Alexander Cowtan, Silas Dilkes, Ross Duncan, Alexandre Krajenbrink, Will
  Simmons, and Seyon Sivarajah.
\newblock On the qubit routing problem.
\newblock {\em arXiv preprint arXiv:1902.08091}, 2019.

\bibitem{shi2020resource}
Yunong Shi, Pranav Gokhale, Prakash Murali, Jonathan~M Baker, Casey Duckering,
  Yongshan Ding, Natalie~C Brown, Christopher Chamberland, Ali Javadi-Abhari,
  Andrew~W Cross, et~al.
\newblock Resource-efficient quantum computing by breaking abstractions.
\newblock {\em Proceedings of the IEEE}, 108(8):1353--1370, 2020.

\bibitem{zhang2020context}
Yu~Zhang, Haowei Deng, and Quanxi Li.
\newblock Context-sensitive and duration-aware qubit mapping for various {NISQ}
  devices.
\newblock {\em arXiv preprint arXiv:2001.06887}, 2020.

\bibitem{tomesh2020coreset}
Teague Tomesh, Pranav Gokhale, Eric~R Anschuetz, and Frederic~T Chong.
\newblock Coreset clustering on small quantum computers.
\newblock {\em arXiv preprint arXiv:2004.14970}, 2020.

\bibitem{farhi2020quantum}
Edward Farhi, David Gamarnik, and Sam Gutmann.
\newblock The quantum approximate optimization algorithm needs to see the whole
  graph: A typical case.
\newblock {\em arXiv preprint arXiv:2004.09002}, 2020.

\bibitem{bravyi2019obstacles}
Sergey Bravyi, Alexander Kliesch, Robert Koenig, and Eugene Tang.
\newblock Obstacles to state preparation and variational optimization from
  symmetry protection.
\newblock {\em arXiv preprint arXiv:1910.08980}, 2019.

\bibitem{amdahl1967validity}
Gene~M Amdahl.
\newblock Validity of the single processor approach to achieving large scale
  computing capabilities.
\newblock In {\em Proceedings of the April 18-20, 1967, spring joint computer
  conference}, pages 483--485, 1967.

\bibitem{wootters1982single}
William~K Wootters and Wojciech~H Zurek.
\newblock A single quantum cannot be cloned.
\newblock {\em Nature}, 299(5886):802--803, 1982.

\bibitem{Qiskit}
H{\'e}ctor Abraham, AduOffei, Ismail~Yunus Akhalwaya, Gadi Aleksandrowicz,
  Thomas Alexander, Eli Arbel, Abraham Asfaw, et~al.
\newblock Qiskit: An open-source framework for quantum computing, 2019.
\newblock \url{https://github.com/Qiskit/qiskit}.

\bibitem{cirq}
Cirq, a python framework for creating, editing, and invoking noisy intermediate
  scale quantum {(NISQ)} circuits.
\newblock \url{https://github.com/quantumlib/Cirq}.

\bibitem{svore2018q}
Krysta Svore, Alan Geller, Matthias Troyer, John Azariah, Christopher Granade,
  Bettina Heim, Vadym Kliuchnikov, Mariia Mykhailova, Andres Paz, and Martin
  Roetteler.
\newblock Q\# enabling scalable quantum computing and development with a
  high-level dsl.
\newblock In {\em Proceedings of the Real World Domain Specific Languages
  Workshop 2018}, pages 1--10, 2018.

\bibitem{cross2017open}
Andrew~W Cross, Lev~S Bishop, John~A Smolin, and Jay~M Gambetta.
\newblock Open quantum assembly language.
\newblock {\em arXiv preprint arXiv:1707.03429}, 2017.

\bibitem{mckay2018qiskit}
David~C McKay, Thomas Alexander, Luciano Bello, Michael~J Biercuk, Lev Bishop,
  Jiayin Chen, Jerry~M Chow, Antonio~D C{\'o}rcoles, Daniel Egger, Stefan
  Filipp, et~al.
\newblock Qiskit backend specifications for openqasm and openpulse experiments.
\newblock {\em arXiv preprint arXiv:1809.03452}, 2018.

\bibitem{singhalverified}
Kartik Singhal, Robert Rand, and Michael Hicks.
\newblock Verified translation between low-level quantum languages.
\newblock In {\em Informal Proceedings of the First International Workshop on
  Programming Languages for Quantum Computing (PLanQC '20)}, 2020.

\bibitem{litteken2020updated}
Andrew Litteken, Yung-Ching Fan, Devina Singh, Margaret~R Martonosi, and Fred
  Chong.
\newblock An updated {LLVM}-based quantum research compiler with further
  openqasm support.
\newblock {\em Quantum Science and Technology}, 2020.

\bibitem{iten2019introduction}
Raban Iten, Oliver Reardon-Smith, Luca Mondada, Ethan Redmond, Ravjot~Singh
  Kohli, and Roger Colbeck.
\newblock Introduction to {UniversalQCompiler}.
\newblock {\em arXiv preprint arXiv:1904.01072}, 2019.

\bibitem{li2020qasmbench}
Ang Li and Sriram Krishnamoorthy.
\newblock Qasmbench: A low-level qasm benchmark suite for {NISQ} evaluation and
  simulation.
\newblock {\em arXiv preprint arXiv:2005.13018}, 2020.

\bibitem{alexander2020qiskit}
Thomas Alexander, Naoki Kanazawa, Daniel~J Egger, Lauren Capelluto,
  Christopher~J Wood, Ali Javadi-Abhari, and David McKay.
\newblock Qiskit pulse: Programming quantum computers through the cloud with
  pulses.
\newblock {\em arXiv preprint arXiv:2004.06755}, 2020.

\bibitem{krantz2019quantum}
Philip Krantz, Morten Kjaergaard, Fei Yan, Terry~P Orlando, Simon Gustavsson,
  and William~D Oliver.
\newblock A quantum engineer's guide to superconducting qubits.
\newblock {\em Applied Physics Reviews}, 6(2):021318, 2019.

\bibitem{arute2019quantum}
Frank Arute, Kunal Arya, Ryan Babbush, Dave Bacon, Joseph~C Bardin, Rami
  Barends, Rupak Biswas, Sergio Boixo, Fernando~GSL Brandao, David~A Buell,
  et~al.
\newblock Quantum supremacy using a programmable superconducting processor.
\newblock {\em Nature}, 574(7779):505--510, 2019.

\bibitem{smith2016practical}
Robert~S Smith, Michael~J Curtis, and William~J Zeng.
\newblock A practical quantum instruction set architecture.
\newblock {\em arXiv preprint arXiv:1608.03355}, 2016.

\bibitem{maslov2017basic}
Dmitri Maslov.
\newblock Basic circuit compilation techniques for an ion-trap quantum machine.
\newblock {\em New Journal of Physics}, 19(2):023035, 2017.

\bibitem{sorensen1999quantum}
Anders S{\o}rensen and Klaus M{\o}lmer.
\newblock Quantum computation with ions in thermal motion.
\newblock {\em Physical review letters}, 82(9):1971, 1999.

\bibitem{rueschlikon}
Qiskit Team.
\newblock {IBMQ} device information: {IBM Q} 16 rueschlikon {V1.x.x}.
\newblock
  \url{https://github.com/Qiskit/ibmq-device-information/tree/master/backends/rueschlikon/V1},
  2018.

\bibitem{calibrating_two_qubit_gate}
Qiskit Team.
\newblock Calibrating a two-qubit gate.
\newblock
  \url{https://github.com/Qiskit/qiskit-tutorials/blob/40af3da7aa86ce190d04f147daf46fbc893a1966/qiskit/advanced/ignis/1b_calibrating_a_two_qubit_gate.ipynb},
  2019.

\bibitem{mckay2017efficient}
David~C McKay, Christopher~J Wood, Sarah Sheldon, Jerry~M Chow, and Jay~M
  Gambetta.
\newblock Efficient {Z} gates for quantum computing.
\newblock {\em Physical Review A}, 96(2):022330, 2017.

\bibitem{nielsen2002quantum}
Michael~A. Nielsen and Isaac~L. Chuang.
\newblock {\em Quantum Computation and Quantum Information}.
\newblock Cambridge University Press, USA, 10th edition, 2011.

\bibitem{gokhale2020optimized}
Pranav Gokhale, Ali Javadi-Abhari, Nathan Earnest, Yunong Shi, and Frederic~T
  Chong.
\newblock Optimized quantum compilation for near-term algorithms with
  openpulse.
\newblock {\em arXiv preprint arXiv:2004.11205}, 2020.

\bibitem{cooper1986interprocedural}
Keith~D Cooper, Ken Kennedy, and Linda Torczon.
\newblock Interprocedural optimization: Eliminating unnecessary recompilation.
\newblock In {\em Proceedings of the 1986 SIGPLAN symposium on Compiler
  construction}, pages 58--67, 1986.

\bibitem{hall1991managing}
Mary~Wolcott Hall.
\newblock {\em Managing interprocedural optimization}.
\newblock PhD thesis, 1991.

\bibitem{heckey2015compiler}
Jeff Heckey, Shruti Patil, Ali JavadiAbhari, Adam Holmes, Daniel Kudrow,
  Kenneth~R Brown, Diana Franklin, Frederic~T Chong, and Margaret Martonosi.
\newblock Compiler management of communication and parallelism for quantum
  computation.
\newblock In {\em Proceedings of the Twentieth International Conference on
  Architectural Support for Programming Languages and Operating Systems}, pages
  445--456, 2015.

\bibitem{Note1}
This $R_x(180)$ gate is what accounts for the additional $540 = 3 \times 180$
  degrees of active rotation in Table~\ref {tab:SWAP}.

\bibitem{sheldon2016procedure}
Sarah Sheldon, Easwar Magesan, Jerry~M Chow, and Jay~M Gambetta.
\newblock Procedure for systematically tuning up cross-talk in the
  cross-resonance gate.
\newblock {\em Physical Review A}, 93(6):060302, 2016.

\bibitem{Note2}
The right $R_x(180)$ has a $R_z(90)$ in the way, but can be commuted using the
  identity that $R_x(180)R_z(90) = R_z(-90)R_x(180)$.

\bibitem{superstaq}
SuperstaQ~Development Team.
\newblock {SuperstaQ}: Connecting applications to quantum hardware.
\newblock \url{www.super.tech/about-superstaq}, 2021.

\bibitem{magesan2012efficient}
Easwar Magesan, Jay~M Gambetta, Blake~R Johnson, Colm~A Ryan, Jerry~M Chow,
  Seth~T Merkel, Marcus~P Da~Silva, George~A Keefe, Mary~B Rothwell, Thomas~A
  Ohki, et~al.
\newblock Efficient measurement of quantum gate error by interleaved randomized
  benchmarking.
\newblock {\em Physical review letters}, 109(8):080505, 2012.

\bibitem{gambetta2012characterization}
Jay~M Gambetta, AD~C{\'o}rcoles, Seth~T Merkel, Blake~R Johnson, John~A Smolin,
  Jerry~M Chow, Colm~A Ryan, Chad Rigetti, S~Poletto, Thomas~A Ohki, et~al.
\newblock Characterization of addressability by simultaneous randomized
  benchmarking.
\newblock {\em Physical review letters}, 109(24):240504, 2012.

\bibitem{gambetta19cramming}
Jay Gambetta and Sarah Sheldon.
\newblock Cramming more power into a quantum device.
\newblock
  \url{https://www.ibm.com/blogs/research/2019/03/power-quantum-device/}, 2019.

\bibitem{sundaresan2020reducing}
Neereja Sundaresan, Isaac Lauer, Emily Pritchett, Easwar Magesan, Petar
  Jurcevic, and Jay~M Gambetta.
\newblock Reducing unitary and spectator errors in cross resonance with
  optimized rotary echoes.
\newblock {\em arXiv preprint arXiv:2007.02925}, 2020.

\bibitem{seltman2012approximations}
Howard Seltman.
\newblock Approximations for mean and variance of a ratio.
\newblock \url{https://www.stat.cmu.edu/~hseltman/files/ratio.pdf}, 2012.

\bibitem{Note3}
The exception is the $17 \leftrightarrow 18$ pair. The point estimate for the
  SWAP error reduction is 1.02 which is `only' $0.8\sigma $ better than 1.0
  parity. We attribute this to the fact that $17 \leftrightarrow 18$ pair is
  also the pair with highest absolute error.

\bibitem{bernstein1997quantum}
Ethan Bernstein and Umesh Vazirani.
\newblock Quantum complexity theory.
\newblock {\em SIAM Journal on computing}, 26(5):1411--1473, 1997.

\bibitem{gokhale2019asymptotic}
Pranav Gokhale, Jonathan~M Baker, Casey Duckering, Natalie~C Brown, Kenneth~R
  Brown, and Frederic~T Chong.
\newblock Asymptotic improvements to quantum circuits via qutrits.
\newblock In {\em Proceedings of the 46th International Symposium on Computer
  Architecture}, pages 554--566, 2019.

\bibitem{gokhale2020extending}
Pranav Gokhale, Jonathan~M Baker, Casey Duckering, Frederic~T Chong, Natalie~C
  Brown, and Kenneth~R Brown.
\newblock Extending the frontier of quantum computers with qutrits.
\newblock {\em IEEE Micro}, 40(3):64--72, 2020.

\bibitem{qiskit2020weighted}
Qiskit~Development Team.
\newblock Weighted adder.
\newblock
  \url{https://qiskit.org/documentation/stubs/qiskit.circuit.library.WeightedAdder.html}.

\bibitem{zlokapa2020boundaries}
Alexander Zlokapa, Sergio Boixo, and Daniel Lidar.
\newblock Boundaries of quantum supremacy via random circuit sampling.
\newblock {\em arXiv preprint arXiv:2005.02464}, 2020.

\bibitem{poyatos1997complete}
JF~Poyatos, J~Ignacio Cirac, and Peter Zoller.
\newblock Complete characterization of a quantum process: the two-bit quantum
  gate.
\newblock {\em Physical Review Letters}, 78(2):390, 1997.

\bibitem{chuang1997prescription}
Isaac~L Chuang and Michael~A Nielsen.
\newblock Prescription for experimental determination of the dynamics of a
  quantum black box.
\newblock {\em Journal of Modern Optics}, 44(11-12):2455--2467, 1997.

\bibitem{knill2008randomized}
Emanuel Knill, Dietrich Leibfried, Rolf Reichle, Joe Britton, R~Brad Blakestad,
  John~D Jost, Chris Langer, Roee Ozeri, Signe Seidelin, and David~J Wineland.
\newblock Randomized benchmarking of quantum gates.
\newblock {\em Physical Review A}, 77(1):012307, 2008.

\bibitem{magesan2011scalable}
Easwar Magesan, Jay~M Gambetta, and Joseph Emerson.
\newblock Scalable and robust randomized benchmarking of quantum processes.
\newblock {\em Physical review letters}, 106(18):180504, 2011.

\bibitem{magesan2012characterizing}
Easwar Magesan, Jay~M Gambetta, and Joseph Emerson.
\newblock Characterizing quantum gates via randomized benchmarking.
\newblock {\em Physical Review A}, 85(4):042311, 2012.

\end{thebibliography}

\clearpage
\appendix
\section{Randomized Benchmarking Results} \label{app:rb}

In this Appendix, we experimentally evaluate our Optimized SWAP relative to the Standard SWAP in terms of fidelity. It is tempting to measure the advantage of the Optimized SWAP by preparing a state that has $\ket{1}$ on one qubit and then SWAPping it to a distant qubit. It is true that a good SWAP would ensure that the distant qubit becomes $\ket{1}$---in fact this is one of our application-level benchmarks (Long SWAP) in Section~\ref{subsec:application_level_benchmarks}. However, such a test would overlook potential phase errors---$R_z$ rotations that don't manifest when only working with classical bitstrings. For instance, a SWAP that mistakenly applied an extra $R_z$ gate would pass the Long SWAP test even though the SWAP would be erroneous.

A more complete characterization of the Optimized SWAP would instead require Quantum Process Tomography (QPT) \cite{poyatos1997complete, chuang1997prescription}, which measures every possible dimension of the operation's error. However, QPT is expensive, requiring 144 separate types of experiments. Moreover, QPT is susceptible to State Preparation and Measurement (SPAM) errors.

Instead of trying to measure every dimension of the SWAP's error, we settle for a more modest but nonetheless practical goal: measuring the fidelity of the Optimized SWAP with a procedure known as Randomized Benchmarking (RB) \cite{knill2008randomized, magesan2011scalable, magesan2012characterizing}. Unlike QPT, RB is resilient to SPAM errors. The key idea behind RB is to execute a circuit that would ideally perform a no-op and recover the initial state. However, due to noise, the probability of success (recovering initial state) will exponentially decay from 100\% down to 25\% (for a two-qubit circuit) as the circuit depth increases.

\begin{figure}[h]
\centering
    \includegraphics[width=0.23\textwidth]{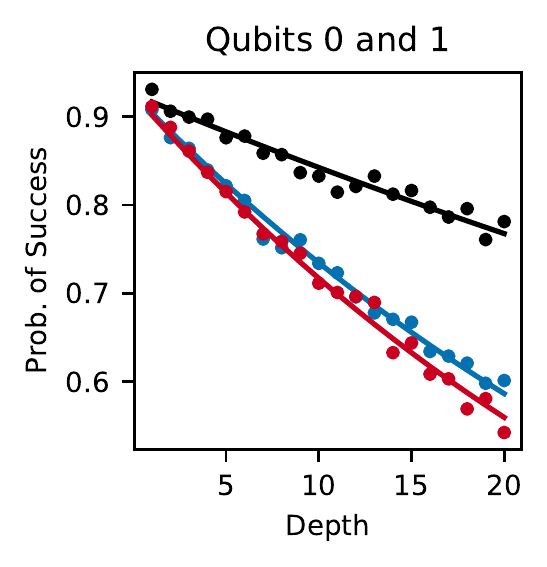}
    \includegraphics[width=0.23\textwidth]{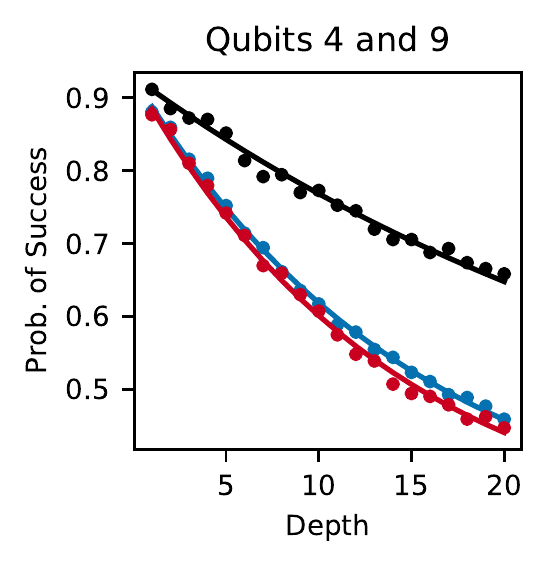}
    \includegraphics[width=0.23\textwidth]{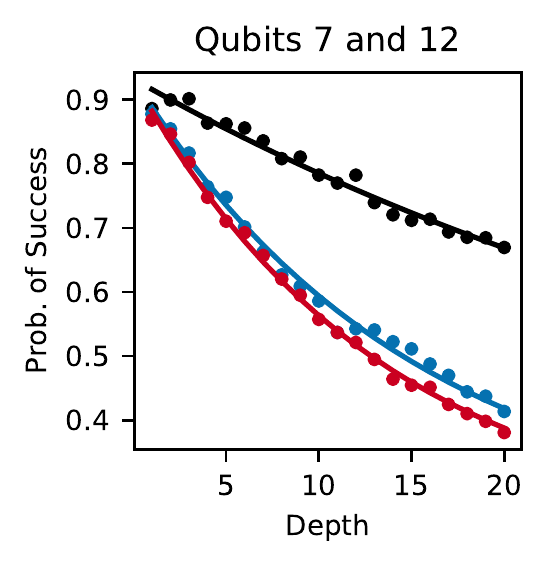}
    \includegraphics[width=0.23\textwidth]{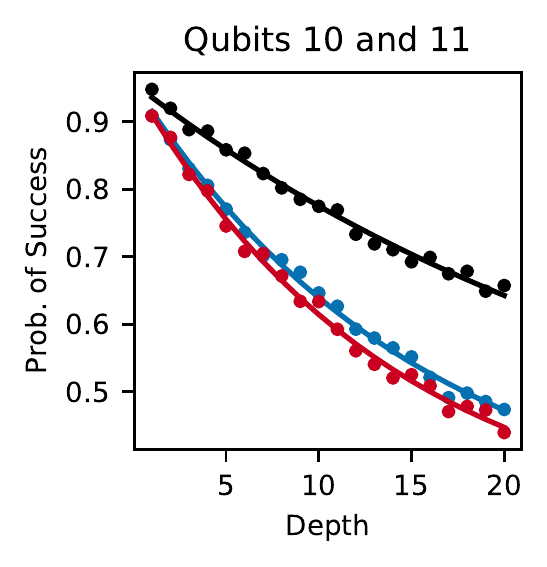}
    \includegraphics[width=0.5\textwidth]{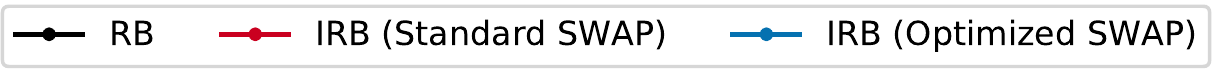}
\caption{(Interleaved) Randomized Benchmarking of the Standard and Optimized SWAPs on IBMQ Johannesburg. We benchmarked all connected qubit pairs; results are plotted from four representative pairs. The black points correspond to ordinary RB. The red and blue points correspond to IRB, which interleaves the RB circuits with Standard or Optimized SWAPs, leading to faster success decay with the depth. Each sequence is fit to an exponential decay. Across all pairs, the Optimized SWAP performs better than the Standard SWAP.}
\label{fig:rb}
\end{figure}

In our case, we want to specifically measure the fidelity of the Optimized SWAP gate, rather than measuring the fidelity of all operations (including single-qubit gates) collectively, as usually performed by RB. We therefore turn to Interleaved Randomized Benchmarking (IRB) \cite{magesan2012efficient}, which serves exactly this purpose. IRB prescribes that we execute both an ordinary RB circuit as well as a modified one that has interleaved SWAP gates. Since the interleaved circuit has more gates, its probability of success decays to 25\% faster than the ordinary RB circuit. The difference in decay rates can be used to calculate the error of the SWAP.

We ran IRB on 22 of the 23 connected qubit pairs on IBMQ Johannesburg. The remaining pair, between qubits 13 and 14, was malfunctioning (reported 100\% error) when we ran our experiments and was therefore excluded. Each of the 22 experiments required $3 \times 20$ circuits evaluated with 8000 shots (repetitions). This totaled 10.5M shots and ran consecutively for several hours. Figure~\ref{fig:rb} shows results from four representative qubit pairs. The blue and red curves correspond to Optimized and Standard SWAPs respectively. Across all pairs, the probability of success (recovering the initial state) decays more slowly for the Optimized SWAP IRB than for the Standard SWAP IRB. This experimentally validates our optimizations.

The black (RB), blue (Optimized SWAP IRB), and red (Standard SWAP IRB) points in Figure~\ref{fig:rb} are all fit to an exponential decay with respective decay parameters $\alpha_\text{RB}, \alpha_\text{Opt}, \text{and } \alpha_\text{Std}$. Given these decay rates, the gate errors of the SWAPs are given by:
\begin{equation}
    \text{error}_\text{Opt} = \frac{3}{4} (1 - \alpha_\text{Opt}/\alpha_\text{RB}), \enspace \text{error}_\text{Std} = \frac{3}{4} (1 - \alpha_\text{Std}/\alpha_\text{RB}) 
\end{equation}
The calculated gate errors are presented in Figure~\ref{fig:swap_error} of the main text. The variances in the exponential decay fit parameters were used to compute the error bars presented in Figure~\ref{fig:swap_error_reduction}.

\newpage
\onecolumngrid
\section{Pulse Schedules} \label{app:pulse_schedules}

\begin{figure*}[ht]
\centering
\begin{minipage}{0.45\textwidth}
\centering
    \includegraphics[width=\columnwidth]{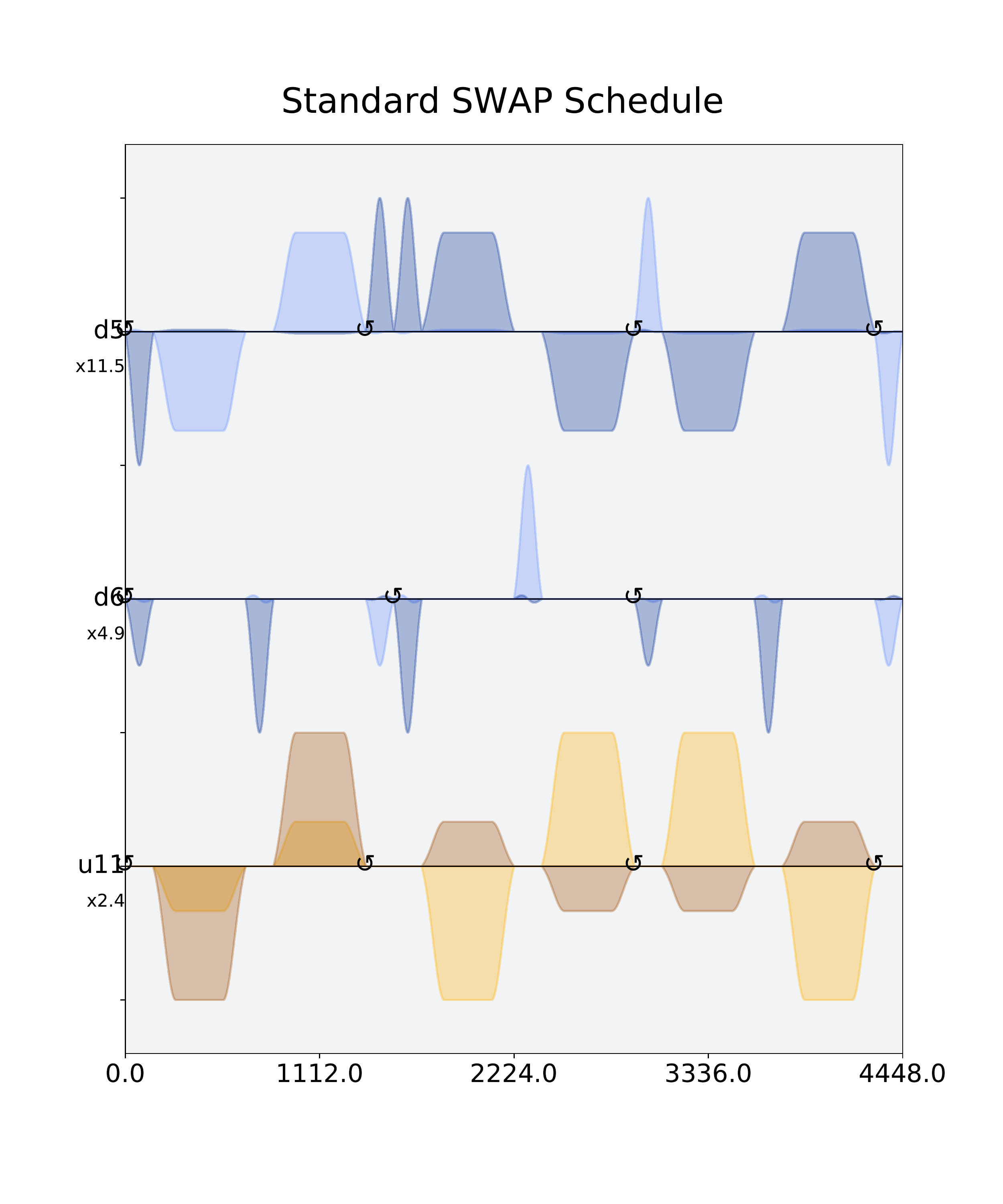}
    \caption{The Standard SWAP pulse schedule between Q5 and Q6 on IBMQ Casablanca. It has a duration of 4448 dt = 988 ns.}
    \label{fig:std_pulse_schedule}
\end{minipage}%
\hspace{5mm}
\begin{minipage}{0.45\textwidth}
\centering
\vspace{14mm}
    \includegraphics[width=\columnwidth]{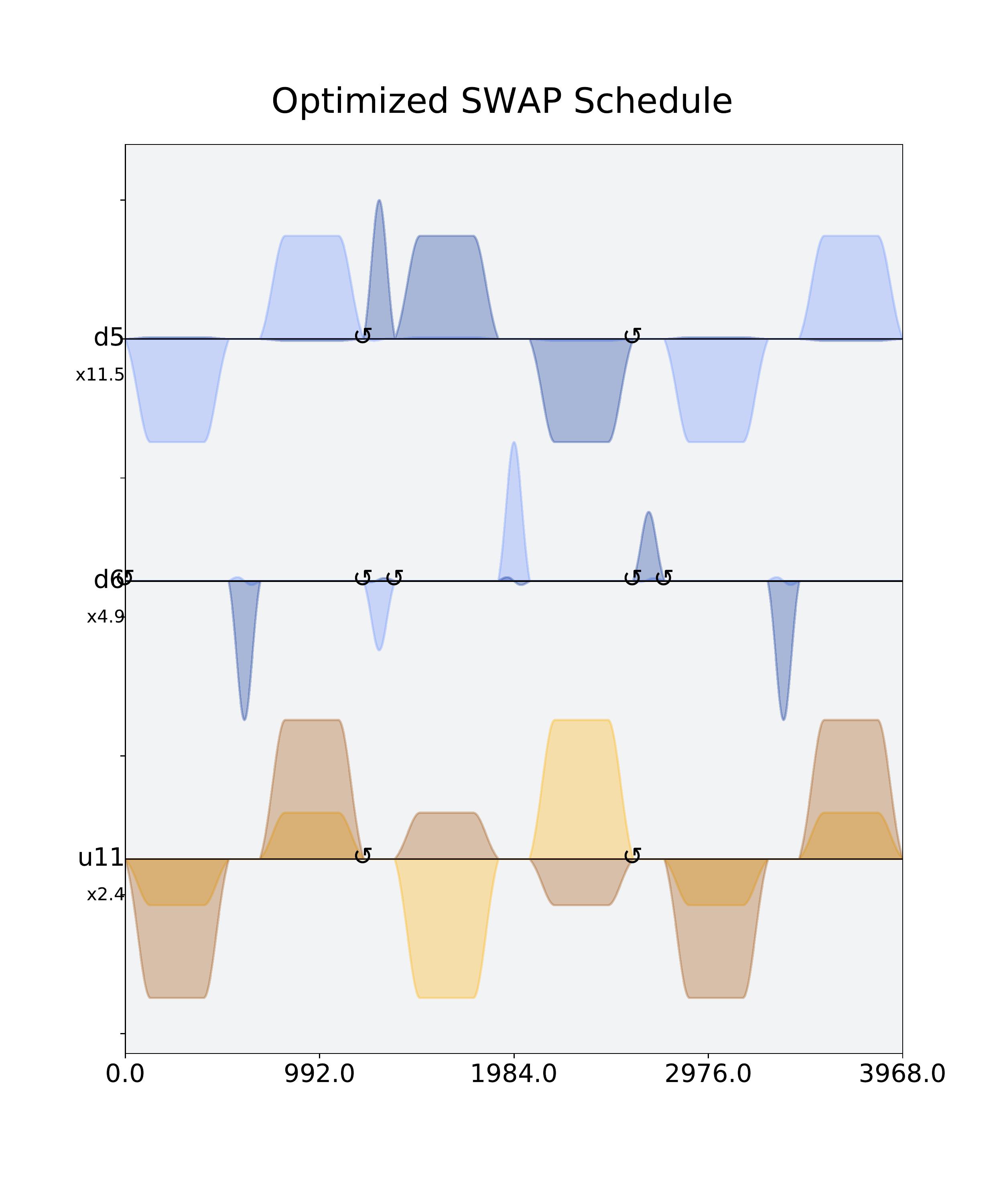}
    \caption{Our Optimized SWAP pulse schedule between Q5 and Q6 on IBMQ Casablanca. It has a duration of 3968 dt = 882 ns, which is 11\% faster than the Standard SWAP. Importantly, our Optimized SWAP is bootstrapped from pre-calibrated native gates, and therefore has zero calibration overhead.}
    \label{fig:opt_pulse_schedule}
\end{minipage}

\begin{minipage}{0.45\textwidth}
\centering
\vspace{10mm}
    $$ \footnotesize
\Qcircuit @C=0.6em @R=1.7em {
& \curvearrowright \qw & \multigate{1}{\downarrow CR_{-+}} & \curvearrowright \qw & \gate{R_x(90)} & \circlearrowright \qw & \multigate{1}{\downarrow CR_{+-}} & \curvearrowleft \qw & \gate{R_x(90)} & \circlearrowright \qw & \multigate{1}{\downarrow CR_{+-}} \\
& \qw & \ghost{\downarrow CR_{-+}} & \curvearrowleft \qw & \gate{R_x(90)} & 
\qw & \ghost{\downarrow CR_{+-}} & \curvearrowleft \qw & \qw & \qw & \ghost{\downarrow CR_{+-}}
}
$$
    \caption{The corresponding circuit view of our Optimized SWAP pulse schedule from Figure~\ref{fig:opt_pulse_schedule}. The circuit is identical to Figure~\ref{fig:optimized_swap}, but with the $R_x(-90)$ gates converted to $R_x(90)$ via virtual $R_z$ gates.}
    \label{fig:optimized_swap_with_x90}
\end{minipage}
      
\end{figure*}
\clearpage
\end{document}